\documentclass[nohyperref]{article}

\usepackage{microtype}
\usepackage{graphicx}
\usepackage{booktabs} 

\usepackage{hyperref}

\usepackage[accepted]{styles/icml2023}

\usepackage{amsmath}
\usepackage{amssymb}
\usepackage{mathtools}
\usepackage{amsthm}
\usepackage{caption}
\usepackage{subcaption}

\usepackage[capitalize,noabbrev]{cleveref}

\theoremstyle{plain}

\theoremstyle{definition}

\theoremstyle{remark}

\usepackage[disable,textsize=tiny]{todonotes}
\usepackage[textsize=tiny]{todonotes}

\usepackage{soul, color}
\newcommand{\Note}[1]{}
\renewcommand{\Note}[1]{#1}  

\icmltitlerunning{Submission and Formatting Instructions for ICML-WCB 2023}

\begin{document}

\twocolumn[
\icmltitle{DiffHopp: A Graph Diffusion Model for Novel Drug Design via Scaffold Hopping}


\icmlsetsymbol{equal}{*}

\begin{icmlauthorlist}
\icmlauthor{Jos Torge}{yyy}
\icmlauthor{Charles Harris}{yyy}
\icmlauthor{Simon V. Mathis}{yyy}
\icmlauthor{Pietro Li\'{o}}{yyy}
\end{icmlauthorlist}

\icmlaffiliation{yyy}{Department of Computer Science and Technology, University of Cambridge}

\icmlcorrespondingauthor{Jos Torge}{jst59@cam.ac.uk}
\icmlcorrespondingauthor{Charles Harris}{cch57@cam.ac.uk}
\icmlcorrespondingauthor{Simon Mathis}{simon.mathis@cl.cam.ac.uk}

\icmlkeywords{Machine Learning, ICML}

\vskip 0.3in
]



\printAffiliationsAndNotice{} 

\begin{abstract}
Scaffold hopping is a drug discovery strategy to generate new chemical entities by modifying the core structure, the \emph{scaffold}, of a known active compound. This approach preserves the essential molecular features of the original scaffold while introducing novel chemical elements or structural features to enhance potency, selectivity, or bioavailability. However, there is currently a lack of generative models specifically tailored for this task, especially in the pocket-conditioned context. In this work, we present DiffHopp, a conditional E(3)-equivariant graph diffusion model tailored for scaffold hopping given a known protein-ligand complex. 
\end{abstract}

\section{Introduction}

\begin{figure}[t!]
\vskip 0.2in
\begin{center}
\centerline{\includegraphics[width=\columnwidth]{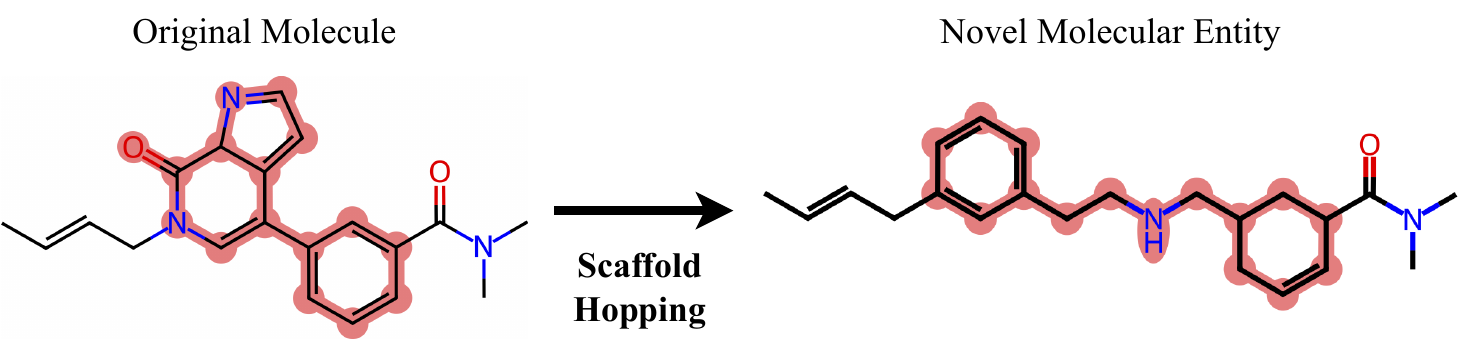}}
\caption{Scaffold hopping: Scaffold atoms highlighted in red. The scaffold holds the functional groups in place for binding. Scaffold hopping refers to interchanging scaffolds while leaving the functional groups unchanged. Compared to inpainting, scaffold hopping uses a fixed definiton of scaffold-vs-rest and in contrast to fragment linking, which links large existing fragments with small molecular bridges, scaffold hopping typically redesigns the majority of a molecule.}
\label{scaffold_hopping}
\end{center}
\vspace{-20pt}
\end{figure}

Scaffold hopping \cite{bohm2004scaffold_hopping} is a widely used strategy in drug discovery that involves modifying the core structure or `scaffold' of a known active compound whilst preserving the functional groups which can be seen as the `business-end' of the molecule which interacts with the target (Figure \ref{scaffold_hopping}). The aim of scaffold hopping is to retain the essential molecular features (also known as pharmacophoric features \cite{yang2010pharmacophore}) of the original scaffold while introducing new chemical elements or structural features that can improve the desired properties, such as potency, selectivity, or bioavailability, whilst designing molecule of novel structure. 

Recently, there has been considerable excitement on the application of deep generative models for many areas within drug discovery \cite{tong2021review1, xie2022review2, isert2023review3, baillif2023review4}, particularly using diffusion models \cite{ho2020ddpm}. There are a number of diffusion models that have been proposed for structure-based drug design \cite{schneuing2022structure}, fragment-linking \cite{igashov2022equivariant} and molecular docking \cite{corso2022diffdock}.

While diffusion models for drug design can in principle be repurposed for scaffold hopping by using an inpainting formulation (appendix~\ref{app:inpainting}), there are no diffusion models specifically designed for scaffold hopping and it is unclear how inpainting with existing models \cite{schneuing2022structure} would compare to a tailored approach. 



In this work, we introduce DiffHopp, an E(3)-equivariant graph diffusion model specifically trained to perform scaffold hopping on known active compounds within protein pockets. Here, we seek to learn the conditional probability distribution of molecular scaffolds given a target pharmacophore. 
In summary, our main contributions are: 
\begin{enumerate}
    \item We \textbf{train a 3D diffusion generative model specifically for the case of scaffold hopping} that is conditioned on whole protein pockets, rather than some desired shape. 
    We further repurpose general pocket conditioned diffusion models \cite{schneuing2022structure} via inpainting (appendix~\ref{app:inpainting}) and observe that specific training for scaffold hopping \textbf{outperforms comparable general models used via inpainting}.
    \item We find that using \textbf{more powerful geometric graph neural networks provides a cure for low connectivity}, a key limitation in current pocket-conditionend molecule generation via diffusion \cite{schneuing2022structure}.
\end{enumerate}


\section{Background and Related Work}

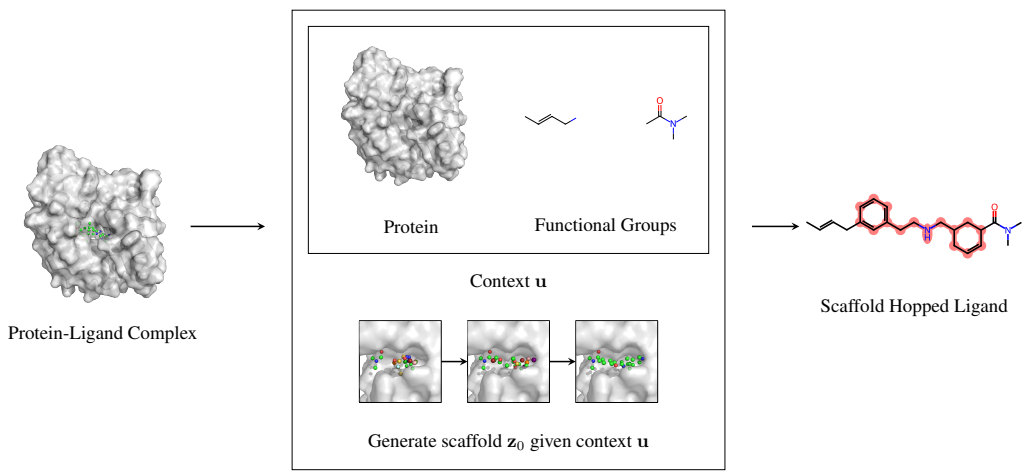
\begin{figure*}[h!]
  \centering
  \tikzstyle{arrow} = [thick,->,>=stealth]
   \resizebox{0.8\textwidth}{!}{
  \begin{tikzpicture}
  	\node (complex) at (-1,0) {\includegraphics[width=3cm]{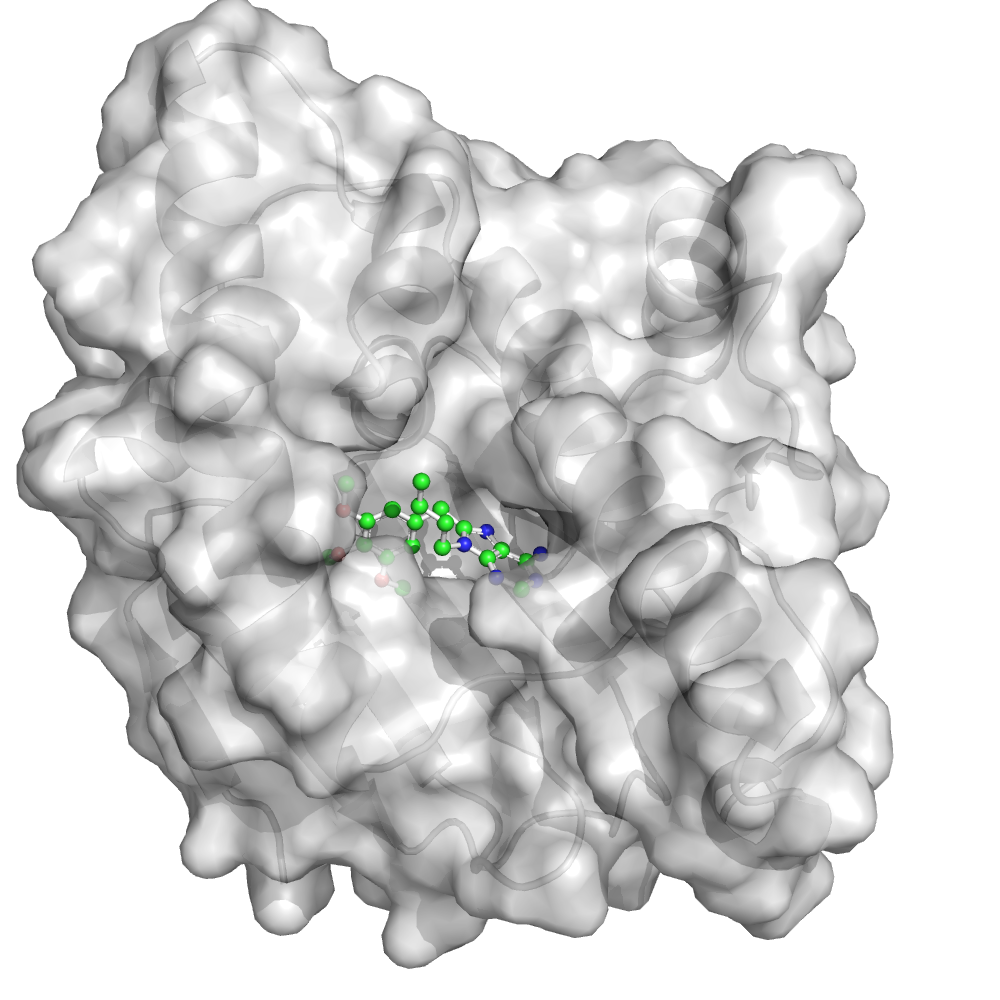}};
  	\node (complex_text) [below of=complex, yshift=-1cm] {Protein-Ligand Complex};
  	
  	\node (protein) at (4.7, 2) {\includegraphics[width=3cm]{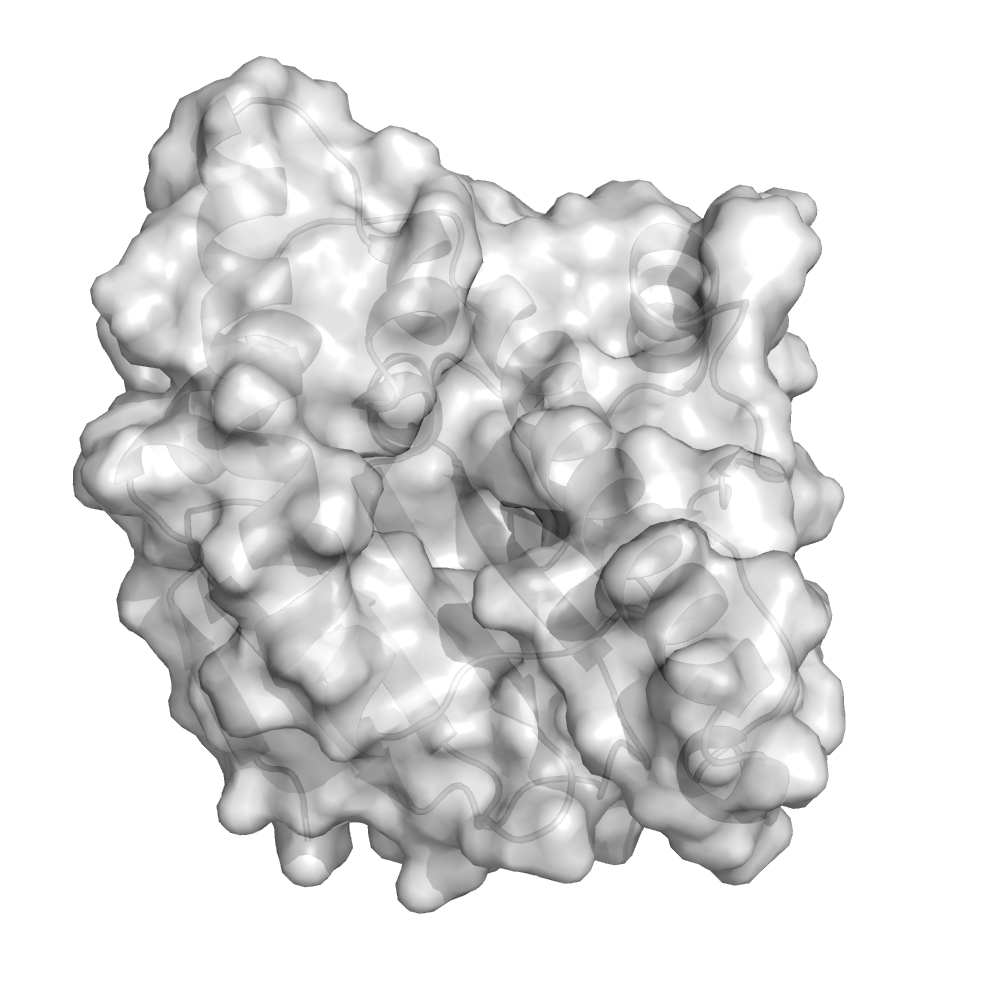}};
  	\node (protein_text) [below of=protein, yshift=-1cm] {Protein};
  	
  	\node (sidechain) at (8.3, 2) {\includegraphics[width=3cm]{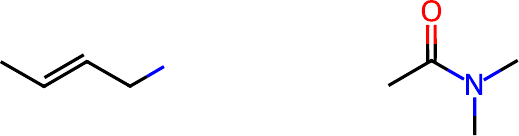}};
  	\node (sidechain_text) [below of=sidechain, yshift=-1cm] {Functional Groups};
  	
  	\draw (2.8,-0.5) rectangle (10.2,3.7);		
  	\node (context) at (6.5,-1) {Context $\mathbf u$};
  	
  	\node (zT) [inner sep=0] at (4.5, -2.5) {\frame{\includegraphics[width=1.5cm]{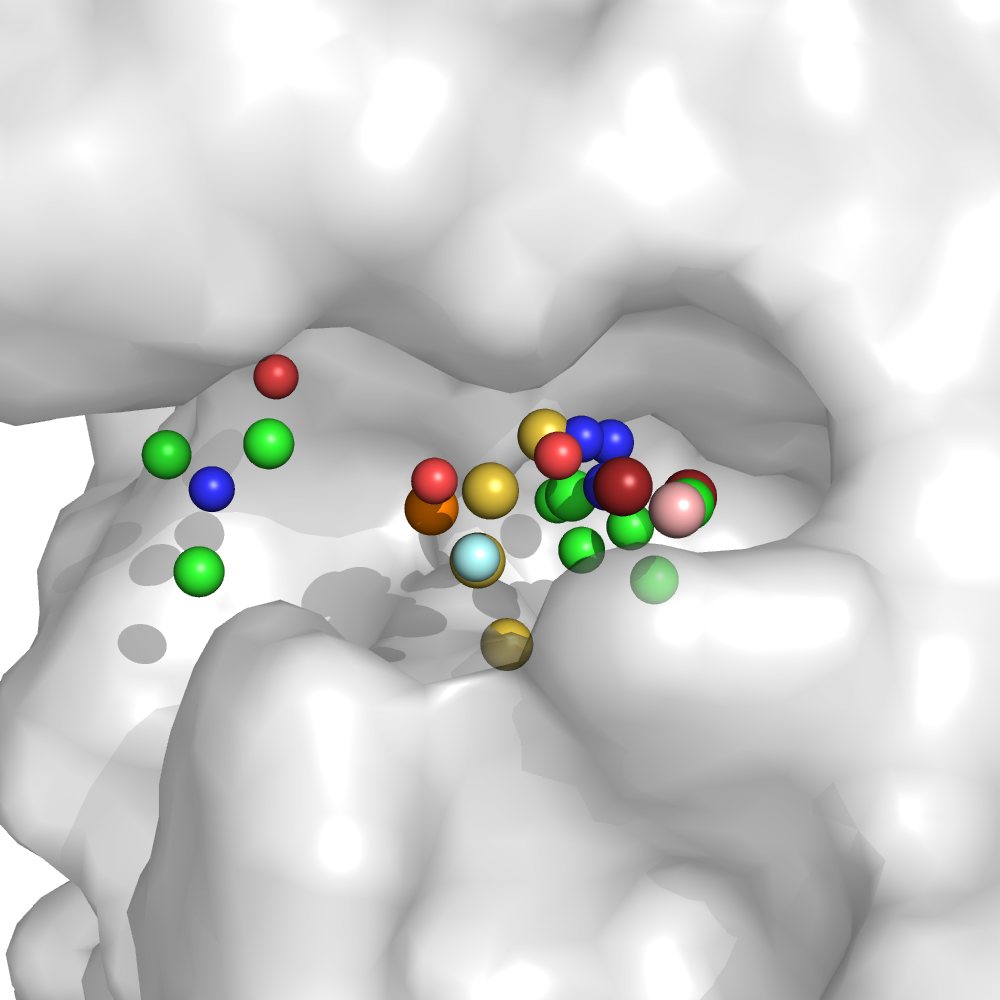}}};
  	\node (zt) [inner sep=0] at (6.5, -2.5) {\frame{\includegraphics[width=1.5cm]{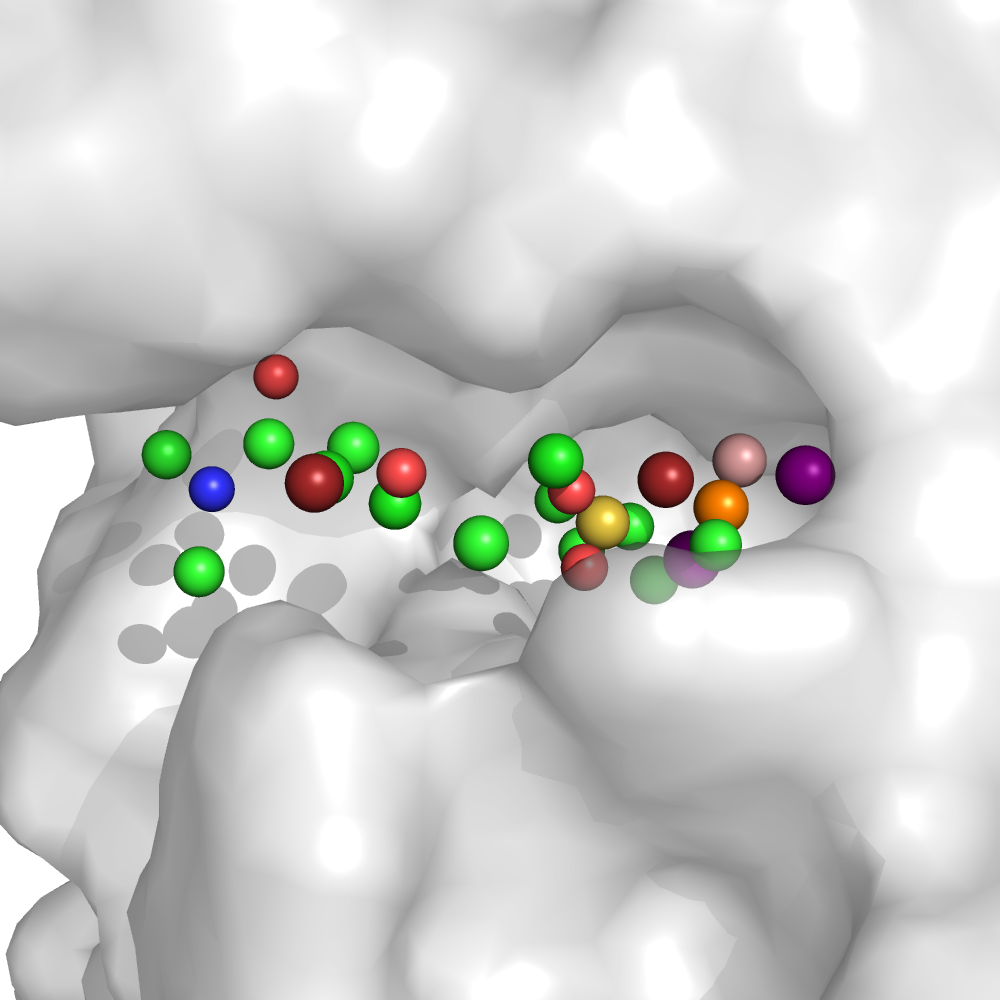}}};
  	\node (z0) [inner sep=0] at (8.5, -2.5) {\frame{\includegraphics[width=1.5cm]{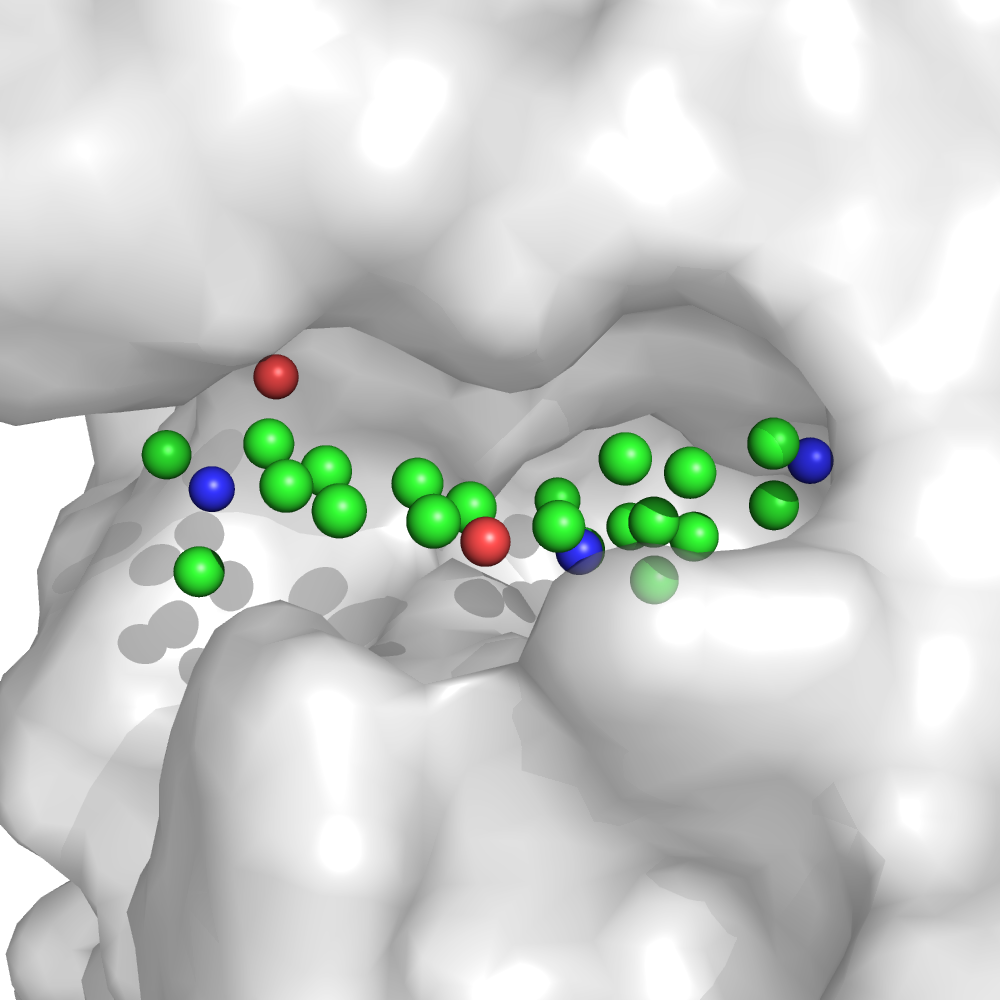}}};
  	\node (diffusion_text) at (6.5, -4){Generate scaffold $\mathbf z_0$ given context $\mathbf u$};
  	
  	\draw (2.5,-4.5) rectangle (10.5,4);
  	\node (scaffoldhop) at (14, 0) {\includegraphics[width=4cm]{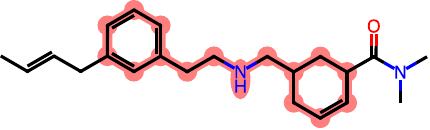}};
  	\node (scaffoldhop_text) [below of=scaffoldhop, yshift=-.5cm] {Scaffold Hopped Ligand};
  	
  	\draw[arrow] (complex) -- (2, 0);
  	\draw[arrow] (11, 0) -- (scaffoldhop);
  	\draw[arrow] (zT) -- (zt);
  	\draw[arrow] (zt) -- (z0);
  	
  \end{tikzpicture}}
  \caption{
  Visual abstract. Given a protein-ligand complex, an equivariant diffusion model is used to sample a scaffold from a scaffold distribution conditioned on functional groups and protein pocket. The resulting scaffold is merged with the functional groups to lead a scaffold hopped ligand.
  }
  \label{fig:visual-abstract}
\end{figure*}

\paragraph{Traditional Scaffold Hopping Methods}

Traditionally, scaffold hopping can be accomplished in different ways \cite{sun2012classification}.
Pharmacophore-based methods define a pharmacophore model which captures common features in known bioactive molecules and then screen large databases for active molecules of novel structure \cite{hessler2010scaffold}. 
Fragment-based methods aim to replace problematic fragments or scaffolds by searching fragment databases based on a simplified chemical similarity \cite{birchall2011reduced}.
However, these non-generative approaches rely on similarity functions, which might not capture the whole spectrum of scaffold relationships. \cite{hu2017recent}

\paragraph{Deep Learning-based Scaffold Hopping}
Early work treated scaffold hopping as sequence translation problem using SMILES \cite{zheng2021smilestranslation}. However, this does not allow reasoning about the 3D chemistry. SQUID \cite{adams2022squid} introduces the first 3D generative model for scaffold hopping, but condition on a desired chemical shape rather than the full receptor chemistry. When used in an inpainting formulation (\citet{lugmayr2022repaint}, appendix~\ref{app:inpainting}), DiffSBDD \cite{schneuing2022structure}, a diffusion model for pocket conditioned ligand generation, can be seen as the closest work to ours. Further, DiffLinker \cite{igashov2022equivariant}, which is trained to generate linkers between molecular fragments using a conditional diffusion model, could in principle be repurposed for scaffold hopping. However, fragment linking typically redesigns small linkers between large fragments, while scaffold hopping normally requires redesigning most of the molecule and is therefore out-of-distribution for the training of fragment linking models.

\paragraph{Diffusion Models} Denoising Diffusion Probabilistic Models (DDPMs) \cite{ho2020ddpm} are a powerful class of generative model used to learn complex probability distributions. In short, DDPMs define a Markovian diffusion process that transforms an observed data distribution into a known prior (typically $\mathcal{N}(\textbf{0}, \textbf{\textit{I}})$). A \textit{score function} (where the score is the \textit{gradient} of the log probability of the underlying density function $\nabla _\textbf{x} \log p(\textbf{x})$) is then learnt to reverse this forward diffusion process, meaning we can sample new data from the tractable prior $\mathcal{N}(\textbf{0}, \textbf{\textit{I}})$ \cite{song2019generative}. 

\section{Methods}

\paragraph{Dataset}
We train our model on 19,378 protein-ligand complexes from PDBBind, filtered for QED $\ge 0.3$ and split as in \citet{corso2022diffdock}. From these complexes, we define the Murko-Bemis scaffold \cite{bemis1996properties} for each ligand using RDKit\footnote{www.rdkit.org}. Here, we treat atoms not in the scaffold as functional groups.

\paragraph{Molecule representation}
All molecules (proteins and ligands) are represented as geometric graphs $\mathcal{G} = \{h, x\}$ with node features $h \in \mathbb{R}^{N\times F}$ and coordinates $x \in \mathbb{R}^{N\times 3}$. Ligands are represented at an atom level, with $h$ being the one-hot encoding of the atom type. For computational efficiency, protein graphs are subset to the pocket region (defined as all atoms within 8~\AA~of the ligand) and are represented at a $C_\alpha$ granularity with node features $h_P$ being the one-hot encoded residue type. Edges within the ligand are fully connected, whereas all protein-ligand and protein-protein edges are drawn with a radius threshold of 5\AA. The edge features between nodes $i$ and $j$ consist of the distance $d_{ij}$ and the normalised direction vector $(\mathbf x_i - \mathbf x_j)/(d_{ij})$.

\paragraph{DiffHopp architecture}
We recast the scaffold hopping problem as learning a conditional probability distribution in 3D, where we wish to construct a new sample scaffold $\textbf{z}_0$ given a molecular context $\textbf{u}$ ( $\textbf{u}$ is the concatenation of the pocket $\textbf{p}$ and functional groups $\textbf{g}$). This is achieved using an equivariant diffusion model $p_\theta (\textbf{z}_0 \mid \textbf{u})$ parameterized using a denoising network $\varepsilon_\theta(\mathbf z_t, t, \mathbf u)$. We parameterize our denoising network $\varepsilon_\theta$ using a diffusion adaptation of the equivariant Geometric Vector Perceptron (GVP) architecture \cite{jing2020learning}. Following previous work \cite{schneuing2022structure, igashov2022equivariant}, we embed all features into a shared feature space using seperate Multi-Layer Perceptrons (MLPs) $\mathbf h_{\textrm{emb}} = [\phi_z(\mathbf h_z), \phi_g(\mathbf h_g), \phi_p(\mathbf h_p)]$ for $z$, $g$ and $p$ respectively. We then perform 7 layers of message passing on the combined pocket-ligand graph to update the hidden node features $h'$ and $x'$. The noise estimator for the scaffold is then taken as $\boldsymbol{\epsilon}_x', \boldsymbol{\epsilon}_h' = \mathbf x'_z, \phi_{\textrm{out}}(\mathbf h'_z)$ with $\phi_{\textrm{out}}$ an MLP to map from embedding space to Gaussian noise.


\paragraph{Training and Sampling}
We follow the DDPM training procedure \cite{ho2020ddpm} outlined in detail in Appendix~\ref{app:sampling} (Algorithm \ref{algo:edm_training}). To ensure equivariance, we employ the zero center of mass trick from previous work \cite{hoogeboom2022equivariant}. We use use $T=500$, AdamW as the optimizer and employ a polynomial variance schedule \cite{hoogeboom2022equivariant} with $s= 10^{-4}$ and all $\alpha_t$ values clipped to a lower bound of $10^{-3}$. We also scale atom features $h$ by 0.25, which was shown in previous work to improve performance empirically \cite{hoogeboom2022equivariant}.
We adapt the simplified noise-prediction objective \cite{schneuing2022structure} into a reweighted loss optimizing atom type and coordinate features individually:
\begin{equation}
	L_{\textrm{reweighted}} = \mathbb E \left[\frac{1}{4} (\lVert\boldsymbol{\epsilon}_x  -\boldsymbol{\epsilon}_x' \rVert^2  + \lVert\boldsymbol{\epsilon}_h  -\boldsymbol{\epsilon}_h' \rVert^2 )\right]
\end{equation} 
where $\left[\boldsymbol{\epsilon}_x, \boldsymbol{\epsilon}_h\right] = \boldsymbol{\epsilon}$ and  $\left[\boldsymbol{\epsilon}_x', \boldsymbol{\epsilon}_h'\right] = \varepsilon_\theta(\mathbf z_t, t, \mathbf u)$ denote the true and predicted noise respectively.
Our sampling procedure follows previous work on equivariant diffusion models \cite{hoogeboom2022equivariant} and is given in Algorithm \ref{algo:edm_sampling} (see appendix). 

\paragraph{Postprocessing}
Following \citet{schneuing2022structure}, we extract the resulting point cloud (fixed functional groups and designed atoms) and convert it into a molecule with bonds using OpenBabel \cite{o2011open}. Molecules are then relaxed using 200 steps of force-field relaxation with UFF \cite{rappe1992uff} to remove clashes.

\section{Experiments}
We set out to answer the following questions: (1) Is a model specifically trained for scaffold hopping much better than a general purpose molecule generation model used with inpainting? (2) What is the effect of using more powerful geometric graph neural networks as denoisers in diffusion for molecule generation?

\paragraph{Evaluation}
To evaluate the quality of generated molecules, we use metrics established in previous work \cite{schneuing2022structure, igashov2022equivariant}. \textbf{Connectivity} measures whether generated molecules are fully connected. \textbf{Diversity} is the average pairwise \textit{Tanimoto-dissimilarity} \cite{bajusz2015tanimoto} between all generated molecules for a pocket. \textbf{Novelty} is the fraction of molecules different from those in the training set. \textbf{QED} \cite{Bickerton2012} is a measure of drug-likeness. \textbf{SA} \cite{ertl2009estimation} estimates ease of synthesis of drug-like molecules. \textbf{Vina Score} is an estimate of binding affinity between ligand and target pocket calculated using the docking software QVina2 \cite{alhossary2015fast}.

\begin{table*}[h]
\centering
\resizebox{\textwidth}{!}{
	\begin{tabular}{lcccccc} 
  Method & Connectivity ($\uparrow$) &  Diversity ($\uparrow$) & Novelty ($\uparrow$) & QED ($\uparrow$) & SA ($\uparrow$) & Vina (kcal/mol, $\downarrow$) \\ 
  \hline
  DiffHopp & \textbf{0.914} {\footnotesize ± 0.28} & 0.592 {\footnotesize ± 0.21} & 0.998 {\footnotesize ± 0.05} & \textbf{0.612} {\footnotesize ± 0.18} & 0.664 {\footnotesize ± 0.13} &  \textbf{-7.883} {\footnotesize ± 1.53}\\
  DiffHopp-EGNN & 0.757 {\footnotesize ± 0.43} & 0.644 {\footnotesize ± 0.17} & \textbf{1.000} {\footnotesize ± 0.02} &  0.514 {\footnotesize ± 0.19} & 0.604 {\footnotesize ± 0.13} &  -7.240 {\footnotesize ± 1.47}\\
  GVP-inpainting & 0.652 {\footnotesize ± 0.48} & \textbf{0.668} {\footnotesize ± 0.18} & 0.997 {\footnotesize± 0.06} & 0.547 {\footnotesize ± 0.20} & \textbf{0.680} {\footnotesize ± 0.11} & -7.552 {\footnotesize ± 1.77} \\ 
  EGNN-inpainting & 0.793 {\footnotesize ± 0.41} & 0.667 {\footnotesize ± 0.18} & 0.999 {\footnotesize± 0.03} &  0.467 {\footnotesize ± 0.20} & 0.644 {\footnotesize ± 0.11} &  -7.163 {\footnotesize ± 1.52} \\
  \hline
  Test set & 1.000 {\footnotesize ± 0.00} & - & 1.000 {\footnotesize ± 0.00} & 0.606 {\footnotesize ± 0.17} &  0.736 {\footnotesize ± 0.12} & -8.767 {\footnotesize ± 1.92} \\
  \end{tabular}
}
\caption{Mean and standard deviation of the common molecular metrics for the molecules from both the test set and the DiffHopp models. Furthermore, results using inpainting on molecule generation models are shown. Best metrics are in bold.}
\label{table:experiment2-metrics}
\end{table*}

\begin{figure}[h]
\vskip 0.2in
\begin{center}
\centerline{\includegraphics[width=0.8\columnwidth]{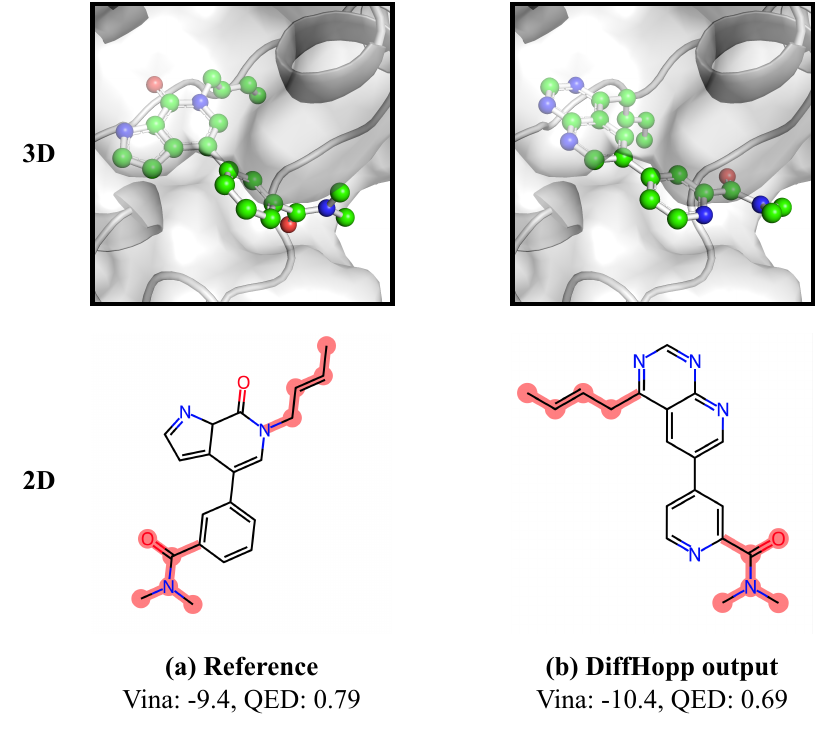}}
\caption{Model output for randomly chosen pocket (PDB:\texttt{6bqd}). Out of 10 generated samples, the best one as measured by Vina score was chosen. Functional groups are highlighted in red. DiffHopp produces more aromatic rings and the molecule has improved binding affinity. }
\label{pymol}
\end{center}
\vskip -0.2in
\end{figure}

\begin{figure}[h]
\vskip 0.2in
\begin{center}
\centerline{\includegraphics[width=\columnwidth]{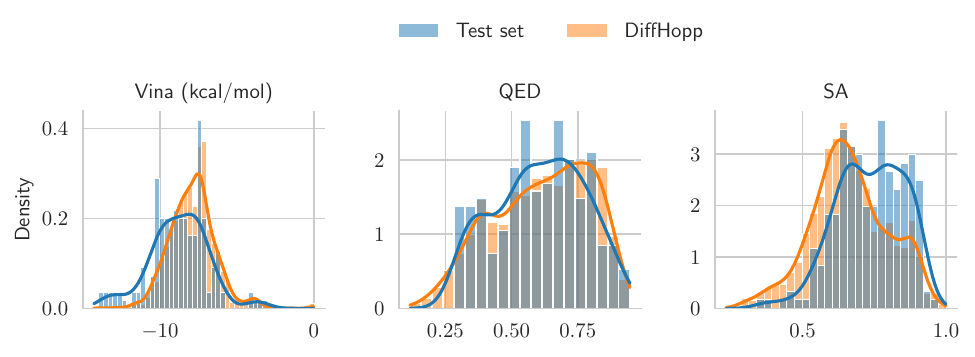}}
\caption{Distribution of selected metrics in the test set and the molecules generated with DiffHopp. The model's samples are comparable in distribution to the test molecules and, while slightly less often than the test set, the model occasionally generates high scoring samples.}
\label{metrics}
\end{center}
\vskip -0.3in
\end{figure}

\paragraph{Scaffold-hopping results}
The main quantitative results for scaffold hopping are presented in Table \ref{table:experiment2-metrics} with distributions of key metrics in Figure \ref{metrics}. Our generated molecules have relatively high chemical diversity, despite the functional groups being fixed in all samples. This indicates that our model can produce molecules of high scaffold/structural diversity. Our mean Vina score of -7.883 is impressive when considering that we often perform drastic topological changes and that the molecules in PDBBind are biased towards high affinity molecules. QED and SA scores are also competitive when compared to the test set and previous work with mean scores of 0.612 \cite{schneuing2022structure} and 0.664 \cite{adams2022squid}. A graphical example of a DiffHopp output is provided in Figure \ref{pymol} for a random target in the test set (PDB:\texttt{6bqd}) \cite{nittinger2019water}. 

We perform dimensionality reduction of the molecular scaffold fingerprints generated by DiffHopp versus the training dataset (Appendix Figure \ref{pca}) and observe that DiffHopp has successfully learned to generate diverse scaffolds that match the training set for diversity chemotypes.

\vskip -0.3in

\paragraph{Comparison to scaffold-hopping via inpainting}
We compared DiffHopp to a general DiffSBDD-like \cite{schneuing2022structure} molecule generation model, trained with the same denoiser. Unlike DiffHopp, this inpainting model was trained without providing functional groups of a ligand as context. To sample scaffolds we fix the functional groups and perform sampling with inpainting (details in App.~\ref{app:inpainting}). DiffHopp showcases clearly superior performance for connectivity, QED, and Vina scores than the inpainting model, while matching its performance in other metrics, barring diversity (Table~\ref{table:experiment2-metrics}). Consequently, our findings affirm that a custom scaffold-hopping model outperforms a repurposed general model via inpainting. The price to pay for the extra performance is the rigid definition of scaffold vs rest, which has to be chosen before training.

\vskip -0.3in

\paragraph{Ablation study of more powerful denoiser}
To test the effect of the more powerful GVP-denoiser, we conducted an ablation study (see Table \ref{table:experiment2-metrics}), replacing the DiffHopp GVP-encoder with an E(3)-Equivariant Graph Neural Network (EGNN) \cite{satorras2021n}. Both models were tuned through extensive hyperparameter optimization.

Our ablation shows that switching from EGNN to GVP significantly improved connectivity, addressing a common problem in EGNN-based works \cite{schneuing2022structure}, where low molecular connectivity due to small coordinate errors causes bond omissions in postprocessing.
We believe this improvement is because GVP is a more expressive model \cite{joshi2023expressive}, which in contrast to EGNN can also reason about angles. This was also reflected in training through reduced coordinate loss compared to the EGNN ablation (Appendix Fig.~\ref{losses}). Additionally, mean QED and Vina scores improved markedly, underlining GVP's superiority. 

\vskip -0.3in



\paragraph{Limitations}

Full atom representations were shown to improve pocket-conditioned diffusion modeling \cite{schneuing2022structure}, however, training such was beyond the computational budget of this project. Future work could investigate whether full-atom representations allow generated scaffolds to better mediate protein-ligand interactions. A related issue is our definition of functional groups as any atom not in the scaffold, which may not capture key pharmacophoric properties contained in the original scaffold (e.g. oxygen bound to the ring in Figure \ref{pymol}).


Whilst the exact size of the medicinally relevant scaffold shape is uncertain, \citet{hu2010global} found that for the majority of targets, between 5-49 structurally distinct scaffolds are available in public databases. Further work will analyse whether DiffHopp is able to enrich the  diversity of chemotypes available for targeting a given protein and whether it generalises to scaffolds beyond those seem in the training set.



\vskip -0.3in

\section{Conclusion}

In this work, we have demonstrated that DiffHopp, an equivariant graph diffusion model, is highly capable of performing the medicinally important task of scaffold hopping to design molecules of potent activity whilst generating novel structures. We found that it outperforms generalist molecule diffusion models used via inpainting and that the expressivity of the denoiser correlated directly with high molecular connectivity. We would thus recommend future work to use more expressive architectures such as GVP. Code will be made available upon acceptance. Code for this work is available at \href{www.github.com/jostorge/diffusion-hopping}{www.github.com/jostorge/diffusion-hopping}.

\section*{Acknowledgements}

The authors would like to thanks Arne Schneuing for his insightful discussions that contributed to this work. SVM was supported by the UKRI Centre for Doctoral Training in Application of Artificial Intelligence to the study of Environmental Risks (EP/S022961/1).

\newpage

\bibliography{bibliography}
\bibliographystyle{styles/icml2023}

\newpage
\appendix
\onecolumn

\section{Training and sampling algorithm}
\label{app:sampling}
We use the same training and sampling algorithms as in \citet{hoogeboom2022equivariant} and \citet{schneuing2022structure}, which is a slight adaptation of the original DDPM sampling \cite{ho2020ddpm}. The main difference to \citet{schneuing2022structure} is the separation of coordinate and atom-type loss in the training.

\begin{algorithm}[h]
\centering
\caption{Training algorithm}
\label{algo:edm_training}
\begin{algorithmic}
\REPEAT
\STATE Sample $\mathbf z_0, \mathbf u$ from training data
\STATE Subtract center of gravity of $\mathbf z_0$ from $\mathbf z_0, \mathbf u$
\STATE Sample $t \sim \mathcal U(0,.., T), \boldsymbol{\epsilon}_x \sim \mathcal N(\mathbf 0, \mathbf I), \boldsymbol{\epsilon}_h \sim \mathcal N(\mathbf 0, \mathbf I)$
\STATE Subtract center of gravity from $\boldsymbol{\epsilon}_x$
\STATE $\boldsymbol{\epsilon} \gets \left[\boldsymbol{\epsilon}_x, \boldsymbol{\epsilon}_h\right]$
\STATE $\mathbf z_t \gets \sqrt{\bar\alpha_t}\mathbf z_0 + \sqrt{1 - \bar\alpha_t}\boldsymbol{\epsilon}$  
\STATE $\left[\boldsymbol{\epsilon}_x', \boldsymbol{\epsilon}_h' \right] \gets \varepsilon_\theta(\mathbf z_t, t, \mathbf u)$
\STATE Take gradient descent step on  $\nabla_\theta (\frac{1}{4}(\lVert\boldsymbol{\epsilon}_x - \boldsymbol{\epsilon}_x' \rVert^2  + \lVert\boldsymbol{\epsilon}_h  -\boldsymbol{\epsilon}_h' \rVert^2 ))$
\UNTIL{convergence}
\end{algorithmic}
\end{algorithm}

\begin{algorithm}[h]
\caption{Sampling algorithm}
\label{algo:edm_sampling}
\begin{algorithmic}
\REQUIRE context $\mathbf u$
\STATE Sample $\mathbf z_T \sim \mathcal N (\mathbf 0, \mathbf I)$
\FOR{$t$ in $T, T - 1, \dots, 1$}
\IF{$t > 1$}
\STATE Sample $\boldsymbol{\epsilon}_x \sim \mathcal N(\mathbf 0, \mathbf I), \boldsymbol{\epsilon}_h \sim \mathcal N(\mathbf 0, \mathbf I)$
\STATE Subtract center of gravity from $\boldsymbol{\epsilon}_x$
\STATE $\boldsymbol{\epsilon} \gets \left[\boldsymbol{\epsilon}_x, \boldsymbol{\epsilon}_h\right]$
\ELSE
\STATE $\boldsymbol{\epsilon} \gets \mathbf 0$
\ENDIF
\STATE $\mathbf z_{t-1} \gets \frac{1}{\sqrt{\alpha_t}}(\mathbf z_t - \frac{\beta_t}{\sqrt{1 - \bar\alpha_t}}\varepsilon_\theta(\mathbf z_t, t, \mathbf u)) + \sigma_t \boldsymbol{\epsilon}$ 
\ENDFOR
\RETURN{$\mathbf z_0$}
\end{algorithmic}
\end{algorithm}

\section{Loss curves}
\begin{figure}[ht]
\vskip 0.2in
\begin{center}
\centerline{\includegraphics[width=0.7\columnwidth]{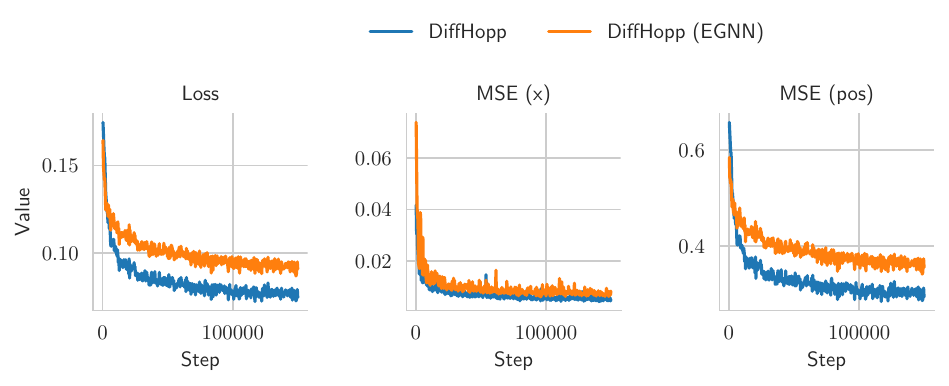}}
\caption{DiffHopp validation losses. Left: total reweighted loss. Middle: node features MSE. Right: Node coordinate MSE.}
\label{losses}
\end{center}
\end{figure}

\section{Performing scaffold-hopping with inpainting}
\label{app:inpainting}
Another approach to scaffold generation is via inpainting: \citet{lugmayr2022repaint} introduce an inpainting method for existing diffusion models to condition their output on known parts. They demonstrate the potential and applicability of the technique by using diffusion models pre-trained for image generation to inpaint images - filling in missing regions. 

It is possible to view the scaffold hopping problem as an inpainting task - using a model trained on de-novo ligand generation, it is possible to consider the scaffold as a missing region while providing the known functional groups of the molecule. 

\begin{figure}
	\centering
	\tikzstyle{arrow} = [thick,->,>=stealth]
	\tikzstyle{op1} = [rectangle, 
  	inner sep=2mm,
	text centered, 
	draw=black, 
	fill=white,
	]
	\resizebox{\textwidth}{!}{
	\begin{tikzpicture}
		\node (side-chain) at (-1,2) {Functional Groups $\mathbf z_0^{\textrm{known}}$};
		\filldraw[blue] (side-chain)+(0.2,0.5) circle (3pt) ;	
		\filldraw[blue] (side-chain)+(-0.3,0.7) circle (3pt) ;	
		\filldraw[blue] (side-chain)+(0.4,0.8) circle (3pt) ;
		\filldraw[blue] (side-chain)+(-0.4,1) circle (3pt) ;	
		\filldraw[blue] (side-chain)+(0,1.3) circle (3pt) ;	
	
		\node (scaffold) at (-1,0) {Scaffold $\mathbf z_{t-1}^{\textrm{unknown}}$};
		\filldraw[black] (scaffold)+(0,1.3) circle (3pt) ;	
		\filldraw[black] (scaffold)+(-0.1,0.8) circle (3pt) ;	
		\filldraw[black] (scaffold)+(0.3,0.7) circle (3pt) ;
		\filldraw[black] (scaffold)+(0.9,0.9) circle (3pt) ;	
		\filldraw[black] (scaffold)+(-0.95,0.9) circle (3pt) ;	
		
		\node (ligand) at (8,1) {Ligand $\mathbf z_{t-1}$};
		\filldraw[blue] (ligand)+(0.3,0.5) circle (3pt) ;	
		\filldraw[blue] (ligand)+(-0.4,0.7) circle (3pt) ;	
		\filldraw[blue] (ligand)+(0.4,0.8) circle (3pt) ;
		\filldraw[blue] (ligand)+(-0.2,0.9) circle (3pt) ;	
		\filldraw[blue] (ligand)+(0,1.5) circle (3pt) ;	
		\filldraw[black] (ligand)+(0,1.3) circle (3pt) ;	
		\filldraw[black] (ligand)+(-0.1,0.8) circle (3pt) ;	
		\filldraw[black] (ligand)+(0.3,0.7) circle (3pt) ;
		\filldraw[black] (ligand)+(0.9,0.9) circle (3pt) ;	
		\filldraw[black] (ligand)+(-0.95,0.9) circle (3pt) ;	
		
		\node (protein) at (12,2.5) {Protein $\mathbf u$};
		\filldraw[red] (protein)+(-0.2,0.5) circle (3pt) ;	
		\filldraw[red] (protein)+(0.3,0.7) circle (3pt) ;	
		\filldraw[red] (protein)+(0.5,0.8) circle (3pt) ;
		\filldraw[red] (protein)+(0.3,0.9) circle (3pt) ;	
		\filldraw[red] (protein)+(0.4,1.5) circle (3pt) ;	
		\filldraw[red] (protein)+(-0.2,1.3) circle (3pt) ;	
		\filldraw[red] (protein)+(-0.3,0.8) circle (3pt) ;	
		\filldraw[red] (protein)+(0.6,0.7) circle (3pt) ;
		\filldraw[red] (protein)+(1.2,0.9) circle (3pt) ;	
		\filldraw[red] (protein)+(-0.65,0.9) circle (3pt) ;	
		
		\node (ligandafter) at (16,1) {Ligand $\mathbf z_{t}$};
		\filldraw[blue] (ligandafter)+(0.25,0.45) circle (3pt) ;	
		\filldraw[blue] (ligandafter)+(-0.45,0.65) circle (3pt) ;	
		\filldraw[blue] (ligandafter)+(0.35,0.85) circle (3pt) ;
		\filldraw[blue] (ligandafter)+(-0.25,0.85) circle (3pt) ;	
		\filldraw[blue] (ligandafter)+(0,1.6) circle (3pt) ;	
		\filldraw[black] (ligandafter)+(0,1.2) circle (3pt) ;	
		\filldraw[black] (ligandafter)+(-0.05,0.75) circle (3pt) ;	
		\filldraw[black] (ligandafter)+(0.3,0.7) circle (3pt) ;
		\filldraw[black] (ligandafter)+(0.9,0.9) circle (3pt) ;	
		\filldraw[black] (ligandafter)+(-0.95,0.9) circle (3pt) ;	
		
		\node at (3.75,2) (noise) [op1] {\tiny $\mathcal{N}(\sqrt{\bar{\alpha}_{t-1}}\mathbf{z}_0, (1 - \bar{\alpha}_{t-1})\mathbf{I})$};
		
		\node at (12, 1) (estimator) [op1, align=center] {Estimator \\ \textit{(de-novo)}};
		
		\node at (15, -1) (extract) [op1] {Extract scaffold and iterate};
		\draw[arrow] (side-chain) -- (noise) -| (6,1) -- (ligand);
		\draw[arrow] (scaffold) -| (6,1) -- (ligand);
		\draw[arrow] (protein) -- (estimator) -- (ligandafter);
		\draw[arrow] (ligand) -- (estimator) -- (ligandafter);
		\draw[arrow] (ligandafter) -| (18, -1) -- (extract) -| (scaffold);
	\end{tikzpicture}
	}
	
	\caption{An overview of reverse diffusion process with inpainting. The ligand $\mathbf z_{t-1}$ is constructed from known and unknown parts, and then iteratively updated.}
	\label{fig:inpainting-process}
\end{figure}
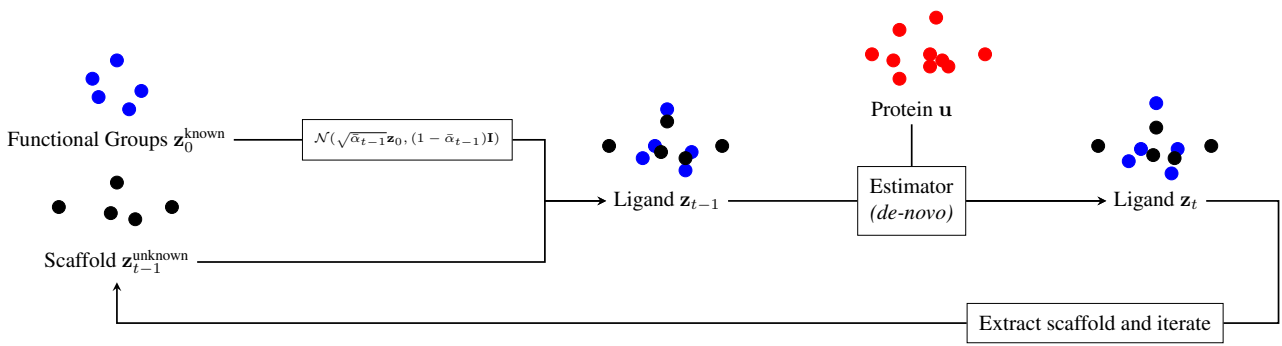

The inpainting method is based on the observation that each step in the reverse diffusion process $p_\theta(\mathbf{z}_{t-1}\vert\mathbf{z}_t)$ depends only on $\mathbf{z}_t$. Thus, it is possible to change $\mathbf{z}_t$ as long as the correct properties of the corresponding distribution are maintained \cite{lugmayr2022repaint}. To create conditioned samples, it is possible to simply enforce the conditioning in the generative process by replacing parts of the predicted $\mathbf{z}_{t-1}$ with the correct $\mathbf{z}_{t-1}^\textrm{known}$.
Formally, given a known $\mathbf z_0$, a current $\mathbf z_{t}$ and a mask indicating the known parts $m$, we can define
\begin{equation}
	\mathbf{z}_{t-1}^{\textrm{known}} \sim \mathcal{N}(\sqrt{\bar{\alpha}_{t-1}} \mathbf{z}_0, (1 - \bar{\alpha}_{t-1})\mathbf{I})
\end{equation} 
\begin{equation}
	\mathbf{z}_{t-1}^{\textrm{unknown}} \sim \mathcal{N}(\boldsymbol{\mu}_\theta(\mathbf{z}_t, t), \boldsymbol{\Sigma}_\theta(\mathbf{z}_t, t))
\end{equation} 
\begin{equation}
	\mathbf{z}_{t-1} = m \odot \mathbf{z}_{t-1}^{\textrm{known}} + (1 - m) \odot \mathbf{z}_{t-1}^{\textrm{unknown}}
\end{equation} 
As the diffusion model attempts to harmonise the input as the diffusion process progresses, this should naturally result in the model generating in-distribution samples with the desired known parts. The process of inpainting is shown in Figure \ref{fig:inpainting-process}.

\begin{figure}[h]
  \centering
  \includegraphics[width=0.6\linewidth]{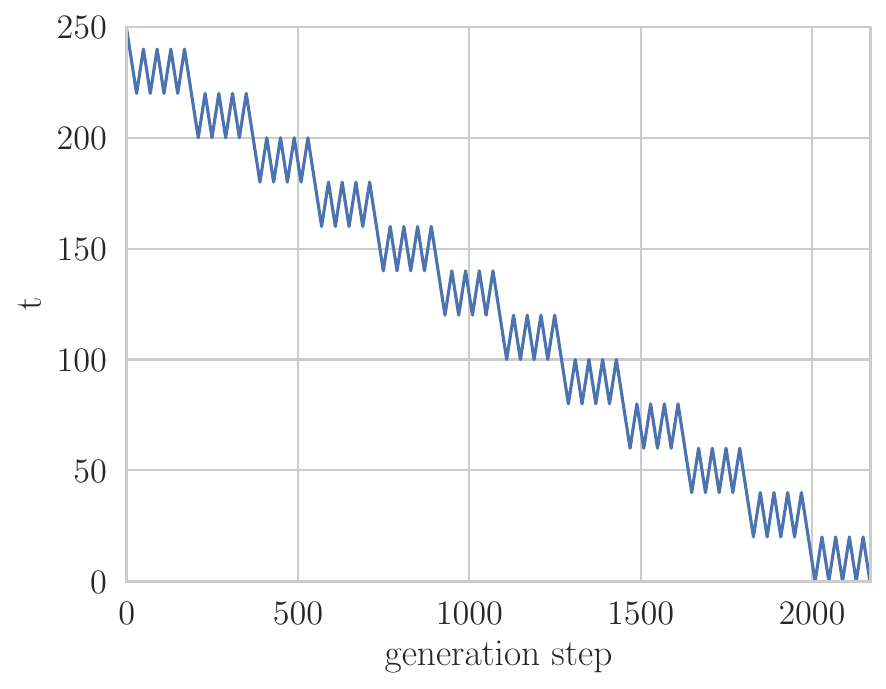}
  \caption{An example repainting schedule with $T = 250$, $r=5$ and $j=20$. Note that resampling always happens $r$ times with length $j$.}
  \label{fig:repaint-schedule}
\end{figure}

However, Lugmayr et al. note that direct application of this method leads to locally harmonised results that struggle to incorporate the global context\footnote{In the paper, the authors describe a case where inpainting the face of a dog leads to a furry texture, not to a face.} \cite{lugmayr2022repaint}. They theorise that the model is limited in how much it can harmonise the sample $\mathbf z_t$ at each step because it does not know about  $\mathbf z_t^{\textrm{known}}$ when making the prediction for $\mathbf z_t^{\textrm{unknown}}$. They compensate for this by not directly following the reverse Markov chain in a linear fashion, but instead moving back and forth in the diffusion process to enable the model to properly incorporate the known parts. This movement is parameterised by $j$ and $r$, where the jump length $j$ indicates the length of each of the $r$ resamples. An example of a repaint schedule is shown in Figure~\ref{fig:repaint-schedule}.

We extend the training procedure to support the training of diffusion models for de-novo ligand generation. In practice, this is done by restricting the context $\mathbf u$ to contain only the protein pocket. Furthermore, $\mathbf z_0$ represents a complete ligand in the training procedure, not just a scaffold. We thus support training models that approximate $p_\theta(\mathbf z_0 | \mathbf u)$, where $\mathbf z_0$ is a ligand and $\mathbf u$ is a protein pocket, similar to the setup of DiffSBDD \cite{schneuing2022structure}). 
\section{Model details}
\label{app:model}
This section details more specific architectural details.  
Following previous work \cite{schneuing2022structure}, the Swish activation function with $\beta = 1$ \cite{ramachandran2017searching}, defined as SiLU \cite{elfwing2018sigmoid}
\begin{equation}
	f(x) = x \cdot \textrm{sigmoid}(x)
\end{equation} 
is used for all non-linearities, except where explicitly detailed otherwise.

\subsection{Encoding and Decoding Functions for the Graph Embedding}
The learnable functions
$\phi_z$, $\phi_g$, $\phi_p$ which encode the respective node features for the shared graph are implemented as Multi Layer Perceptron (MLP). Each function consists of two linear layers, the former mapping from $n$ to $2n$ features and the latter from $2n$ to $m$ features, with one non-linearity in between. $n$ denotes the number of original features and $m$ the size of the joint embedding space (without time appended).

$\phi_{\textrm{out}}$ is a 2-layer MLP with the inverted structure of $\phi_z$.
\subsection{EGNN}
The learnable functions $\phi_e$, $\phi_h$, $\phi_{\textrm{att}}$ and $\phi_x$ are used in each EGNN layer \cite{satorras2021n}. 

Given a value for hidden features $h$, $\phi_e$ is a two-layer MLP mapping from the features of the input to $h$ and then from $h$ to $h$ using two linear layers, with a non-linearity after each of them. The final output is divided by a normalisation factor $C = 100$, to prepare for the sum aggregation.
 
The learnable function $\phi_h$ is another two-layer MLP with a hidden layer size $h$ with a single non-linearity between both layers.

The attention mechanism $\phi_{\textrm{att}}$ is defined as a single linear layer with a single output, followed by a sigmoid function.
 
Finally, the position update $\phi_x = R\tanh(\phi_{x'})$, where $R = 15$ limits the range of movement. $\phi_{x'}$ is a 3-layer MLP with hidden sizes $h$, where the last layer maps to a scalar and has no bias. 

\subsection{GVP-GNN}
\label{sec:gvp-gnn-implementation-details}
This section describes the architecture of the constructed GNN using GVPs \cite{jing2020learning} in more detail. $\sigma_g$ is a sigmoid function in all GVPs mentioned. Unless explicitly stated otherwise, $\sigma$ is the SiLU activation function and $\sigma^+$ the identify function.

The inputs of the GVP are nodes with scalar features $\mathbf s$, representing the embedded input scalars in the graph space, and no vector features. 
The input edges simply consist of a normed direction vector and the distance between the two respective nodes, as detailed in the paper. 

Both edge attributes and node attributes are passed through embedding layers. 
The edges are embedded in a two step process, first normalising the inputs using a layer normalisation \cite{ba2016layer} and then passing them through a GVP with $\sigma, \sigma^+$ being the identify function, which outputs a scalars of hidden size $h/2$ and a single vector.

The nodes are embedded in a similar fashion, however outputting $h/2$ scalars and $h/2$ vectors, leading to $h$ features in total.

The message passing layers are defined similar to the EGNN:

\begin{equation}
\begin{aligned}
\mathbf m_{vw}' &= \phi_e(\mathbf h_v, \mathbf h_w, \mathbf e_{vw}) \\
		\mathbf m_v' &= \sum_{w \in \mathcal N_v}\tilde{e}_{vw}\mathbf m_{vw} \\
		\mathbf h_v' &= \phi_h(\mathbf h_v, \mathbf m_v')  	
		\end{aligned}
\end{equation} 
where $\tilde{e}_{vw} = \phi_\textrm{att}(\mathbf m_{vw})$ is an attention mechanism to learn a soft estimations of the edges, similar to the EGNN.

$\phi_e$ is a composition of three GVPs with hidden sizes $(h/2, h/2)$ and the last one having  $\sigma$ has identity function. $\phi_{\textrm{att}}$ is a single GVP mapping to a single scalar with $\sigma$ being the sigmoid activation function. The final output is normalised by $C = 100$, similar to the EGNN.
 
 $\phi_h(\mathbf h_v, \mathbf m'_v) = \textrm{norm}(\mathbf h_v + \phi_h'(\textrm{norm}(\mathbf h_v + \mathbf m'_v)))$, a residual \cite{he2016deep} architecture with $\phi_h'$ being a composition of two GVPs with $(h/2, h/2)$ has input, hidden and output size. The last layer again has  $\sigma$ as identity function. $\textrm{norm}$ denotes a layer normalisation\cite{ba2016layer}. The norm is not learned for vectors.

\section{Hyperparameter tuning and settings}
\label{app:hyperparam}
We considered the following hyperparameter settings. The best model was chosen by taking
the model with the lowest validation set loss. Hidden Features denote the number of features
for each node between the GNN Message Passing layers. Embedding Size denotes the size of
the node embeddings in the input graph. GNN Layers denotes the number of message passing
layers in the GNN architecture.

\begin{table}[h]
\centering
\resizebox{\textwidth}{!}{
\begin{tabular}{llllll} 
 \textbf{Parameters} & \textbf{Search space} & DiffHopp-EGNN & EGNN-inp. & DiffHopp & GVP-inp. \\ 
  \hline
 \textbf{Static} & \\
 Batch Size &  & 32 \\
 Diffusion steps $T$ & &  500 \\
 Number of steps & & 150000 \\
 Seed & & 1 \\

 \textbf{Tuned} & \\ 
 Attention mechanism & True, False & False & False & True & False\\
 Hidden Features & 32, 64, 128, 256 & 64 & 256 & 256 & 256 \\
 Embedding Size & 32, 64, 128, 256 &  128 & 256 & 64 & 32\\
 Learning rate & 5e-3, 2e-3, 1e-3, 5e-4 & 1e-3 & 1e-3 & 5e-4 & 5e-4 \\
 GNN Layers & 4,5,6,7  & 7 & 7 & 7 & 6 \\ 
\end{tabular}}
\end{table}

\section{Scaffold clustering}

\begin{figure}[h]
\centering
	\centering
    \includegraphics[width=0.5\textwidth]{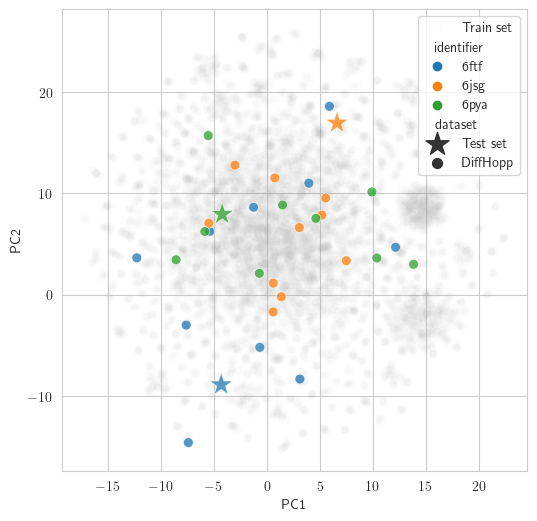}
    \caption{UMAP dimensionality reduction of molecular scaffolds from various sources. Grey: scaffolds in the PDBBinding training dataset. Three molecualrs are chosen at random from the test set (PDB 6FTF, 6JSG and 6PYA in blue, orange and green respectively) and the reduced mapping for the original scaffold (large star) and DiffHopp generated scaffolds (small dot) are shown. DiffHopp is able to generate a large diversity of scaffolds, regardless of the original chemotype specified by the functional groups in the test set molecule.}

\label{pca}
\end{figure}

Figure \ref{pca} shows a dimensionality reduction plot for the scaffolds from the training dataset and those generated using DiffHopp. For each molecule, we select the Murko-Bemis scaffold \cite{bemis1996properties} and calculate its fingerprints using RDKit's topological fingerprints \cite{kearsley1996rdkitfingerprints}, dimensionality reduction is then performed with UMAP \cite{mcinnes2018umap}. We observe that DiffHobb is able to generate diversity scaffolds for a variety of molecules with varying functional groups.

\section{Further examples}
\label{app:samples}
\begin{figure}[h]
	\centering
	PDB:6olx
	
	\vspace{3mm}
	
\begin{subfigure}{0.25\linewidth}
	\centering
		\frame{\includegraphics[width=0.9\textwidth]{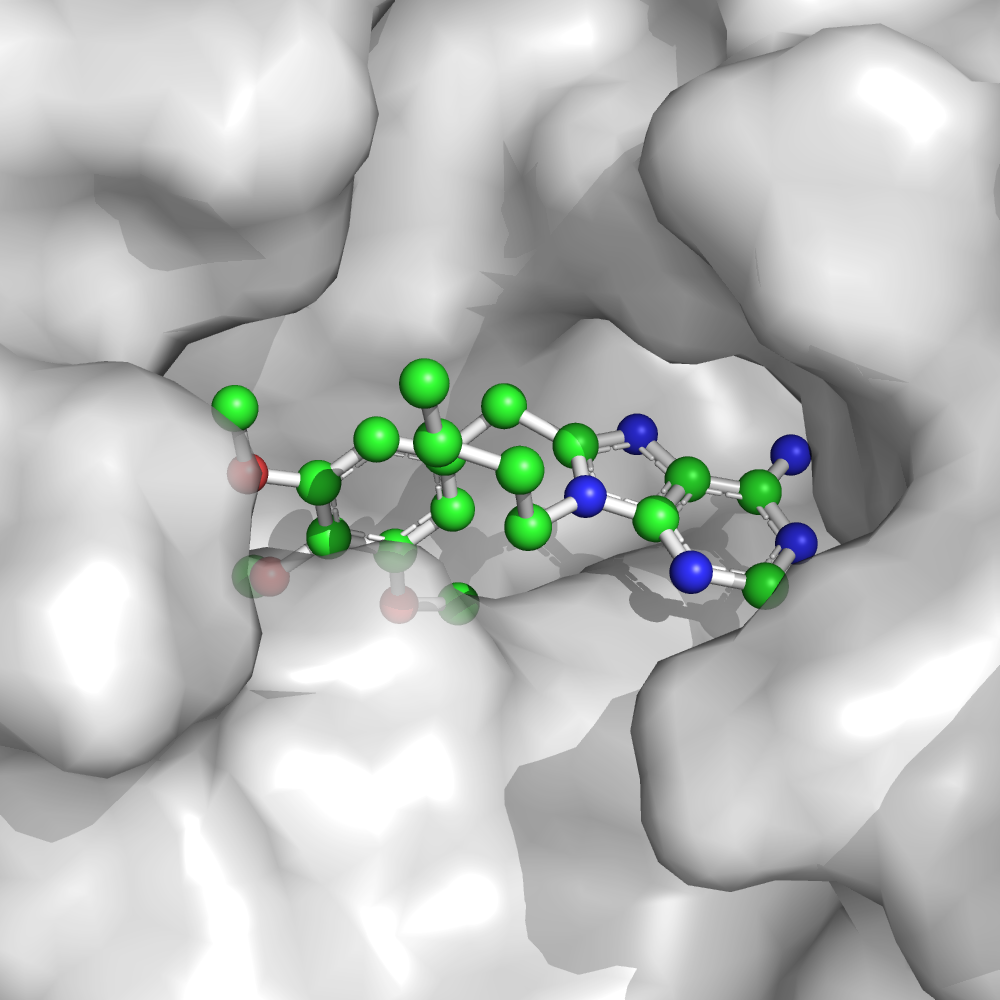}}
	\end{subfigure}
\begin{subfigure}{0.25\textwidth}
\centering
		\frame{\includegraphics[width=0.9\textwidth]{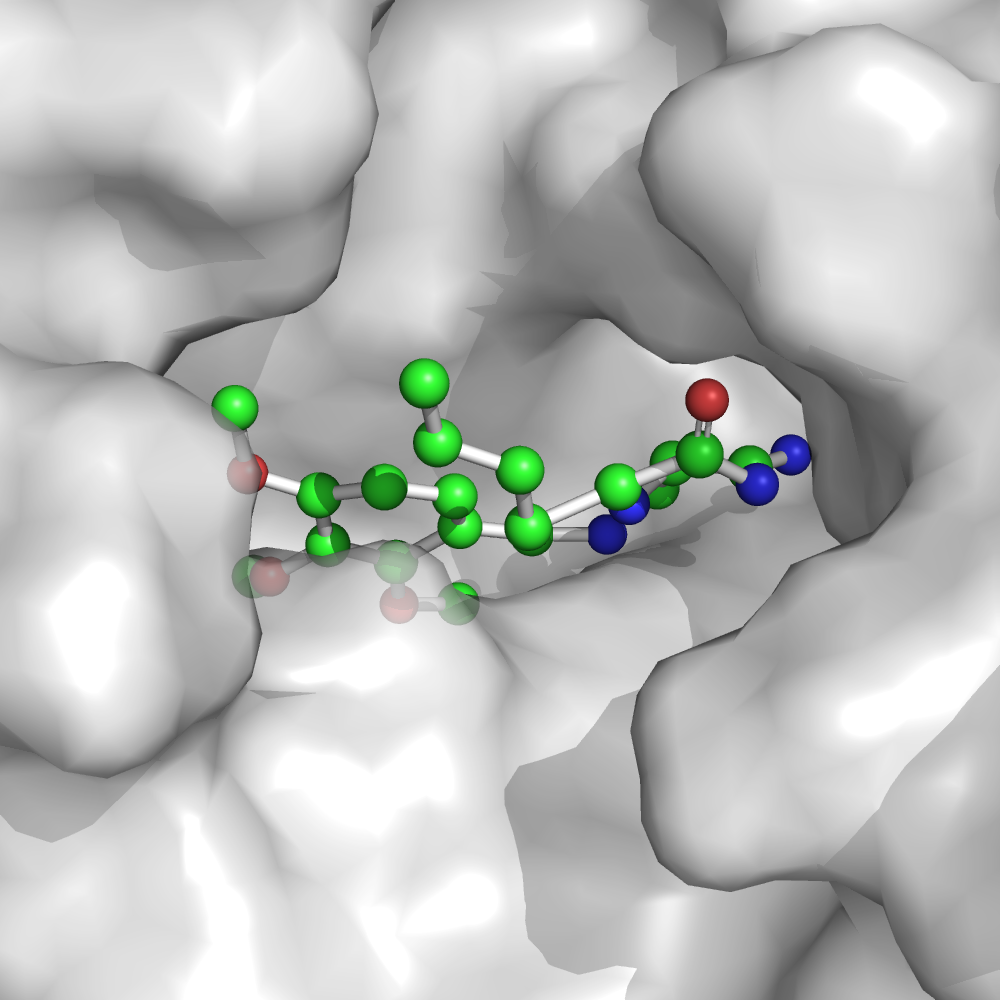}}
\end{subfigure}
\begin{subfigure}{0.25\textwidth}
\centering
		\frame{\includegraphics[width=0.9\textwidth]{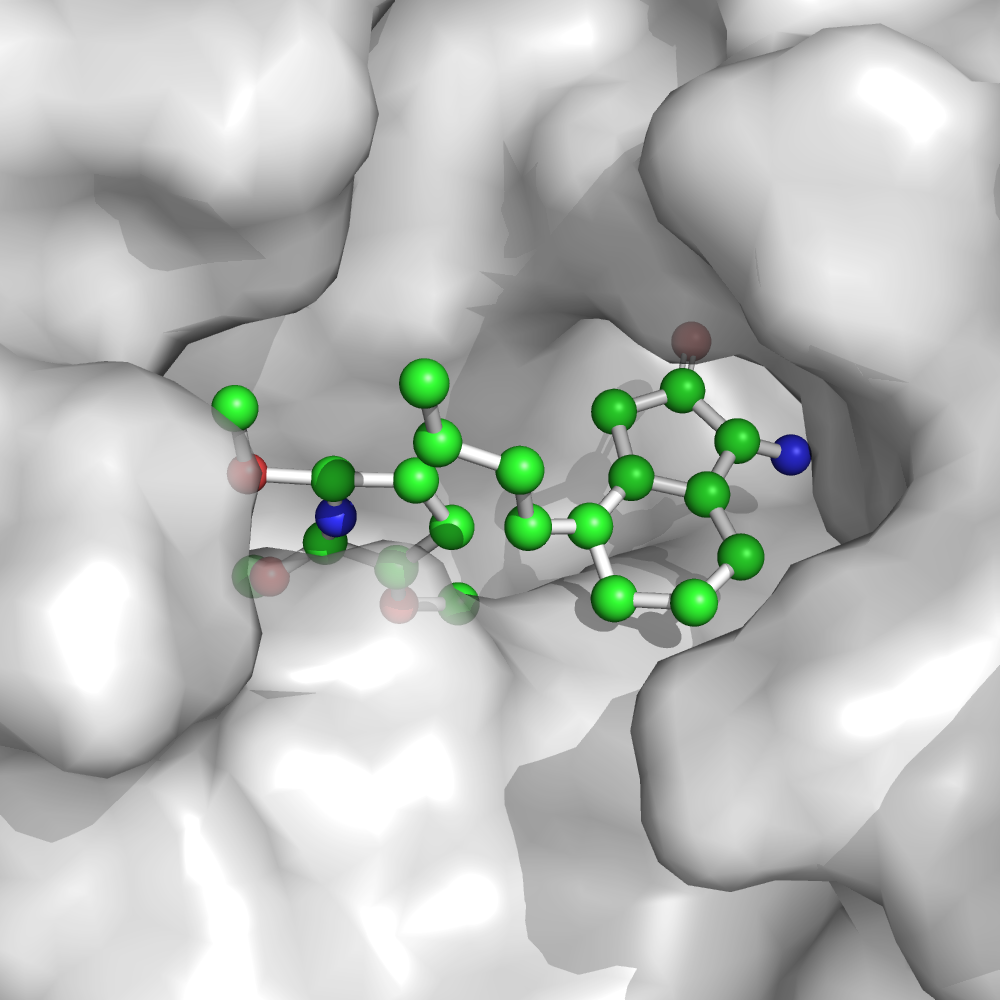}}
\end{subfigure}

\vspace{3mm}

\captionsetup{justification=centering}
\begin{subfigure}{0.25\textwidth}
	\centering
		\includegraphics[width=0.65\textwidth]{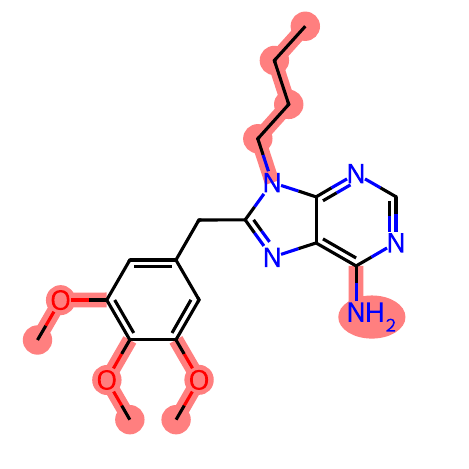}
		\caption{Reference \\ Vina: -8.0, QED: 0.65}
	\end{subfigure}
\begin{subfigure}{0.25\textwidth}
\centering
		\includegraphics[width=0.65\textwidth]{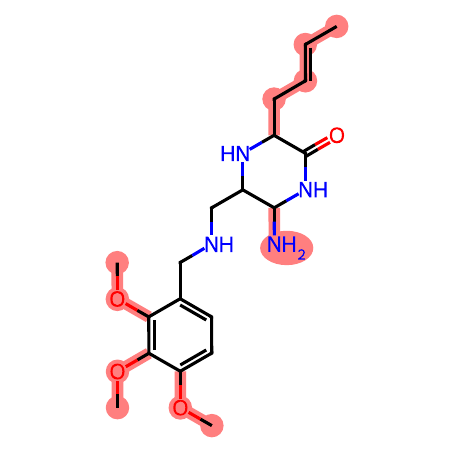}
		\caption{Connected sample \\ Vina: -8.5, QED: 0.69}
\end{subfigure}
\begin{subfigure}{0.25\textwidth}
\centering
		\includegraphics[width=0.65\textwidth]{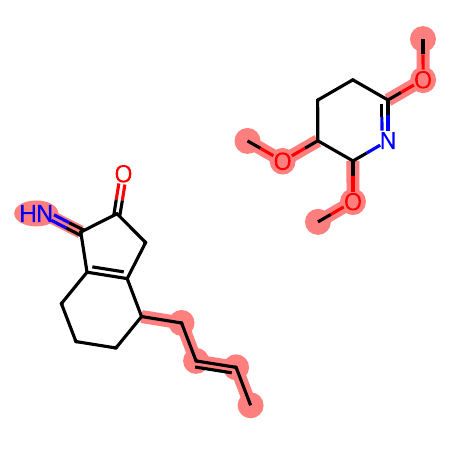}
		\caption{Unconnected sample \\ Vina: -8.4, QED: 0.46}
\end{subfigure}
\captionsetup{justification=default}
\caption{Cherry-picked samples to illustrate capabilities and shortcomings. It can be seen that the model is able to produce scaffolds that perform better on key metrics. However, not all samples are connected. In the chemical structure, functional groups are highlighted in red.}
\label{fig:experiment-suitability-docking-example1}
\end{figure}

To help understand the merits and shortcomings of DiffHopp, we cherrypicked three examples below that illustrate typical successes and shortcomings of DiffHopp (\cref{fig:experiment-suitability-docking-example1}).

\end{document}